\documentclass[aps,pre,nofootinbib,showpacs,twocolumn]{revtex4-1}
\usepackage{natbib}
\usepackage{graphicx}
\usepackage{bm}
\usepackage{gensymb}
\usepackage{color}

\usepackage{color}
\usepackage[normalem]{ulem}

\newcommand {\nn}{\nonumber}
\newcommand {\be}{\begin{equation}}
\newcommand {\ee}{\end{equation}}
\newcommand {\bea}{\begin {eqnarray}}
\newcommand {\eea}{\end{eqnarray}}

\newcommand {\cZ}{{\cal Z}}
\newcommand {\ra}{\rangle}
\newcommand {\la}{\langle}

\begin{document}

\title{Cluster-factorized steady states in finite range processes}
\author{Amit Chatterjee$^1$}
\email{Email : amit.chatterjee@saha.ac.in}
\author{Punyabrata Pradhan$^2$}
\email{Email : punyabrata.pradhan@bose.res.in }
\author{P. K. Mohanty$^{1,3}$ }
\email{Email : pk.mohanty@saha.ac.in}
\affiliation{$^1$Condensed Matter Physics Division, Saha Institute 
of Nuclear Physics, 1/AF Bidhan Nagar, Kolkata 700064, India \\
$^2$Department of Theoretical Sciences, S. N. Bose National Centre 
for Basic Sciences, Block-JD, Sector-III, Salt Lake, Kolkata 700098, India \\ 
$^3$ Max Planck Institute for the Physics of Complex Systems,
N\"othnitzer Stra\ss e 38, 01187 Dresden, Germany}
\date{\today}

\begin{abstract}

We study a class of nonequilibrium lattice models on a ring where particles hop in a particular direction, from a  site to one of its  (say, right) nearest neighbours, with a rate that depends on the
occupation of   all the  neighbouring sites  within a range $R$.   
This finite range process (FRP)  for  $R=0$  reduces to  the well  
known zero-range process (ZRP),   giving rise to a factorized steady 
state (FSS)  for any arbitrary   hop rate. We show  that,  provided 
the hop rates satisfy a specific condition, the steady state  of 
FRP  can be written as a product of  cluster-weight function  
of  $(R+1)$  occupation  variables. We show that, for a large class of cluster-weight functions, the cluster-factorized steady state  admits a finite dimensional transfer-matrix formulation, which helps in  
calculating  the spatial correlation functions and subsystem mass 
distributions  exactly. We also discuss  a criterion  for which the  
FRP undergoes a condensation transition.  

\end{abstract}
%
\maketitle

\section{Introduction}

Driven interacting many-particle systems \cite{Liggett} have been 
of considerable interest in the past decades due to their rich 
transport properties, especially in lower dimensions. The zero 
range process (ZRP), a lattice gas model without any hardcore exclusions, is perhaps the simplest of them, which exhibit nontrivial static and dynamic properties in the steady state. The ZRP was introduced \cite{Spitzer70} as a mathematical model for 
interacting diffusing particles and, since then, has found 
applications in different branches of science \cite{ZRPRev, 
ZRP_Evans}, such as in describing  phase separation criterion in 
driven lattice gases \cite{Criteria}, network re-wiring 
\cite{Network}, statics and dynamics of extended objects 
\cite{ExtendObj,Bijoy}, etc. Interestingly, the ZRP  shows a    
condensation transition for some specific choices of particle 
hop rates for which, when the density becomes larger than a 
critical density $\rho_c$, a macroscopic number of  particles 
accumulate  on a single lattice site - representing
a classical real-space analogue of the Bose-Einstein condensation.  
The ZRP has been generalized to multi-species models 
\cite{Mult},the misanthrope process \cite{Beyond_ZRP}, urn models 
\cite{Urn},the inclusion process \cite{Inclusion} and inhomogeneous hop rates \cite{Inhomo}, etc.

In the ZRP, the particles hop stochastically  to  one of the nearest neighbours with a rate that depends only on the number of particles on the departure site. As a consequence, the ZRP has a 
factorized steady state (FSS), which is amenable  to exact analytic 
studies. However, when the hop rate depends on the neighbouring 
sites,  the steady  state  does  not  factorize in general 
\cite{PFSS_Evans, EvansNew}. In such situations, one may naturally expect a cluster-factorized steady state (CFSS), a straightforward generalization of the factorized steady state (FSS), where the steady 
state weight is a product of cluster-weight functions (see Eq.  
(\ref{eq:CFSS})) of several variables, i.e., the occupation numbers  
at  {\it two}  or more consecutive  sites.

In this paper, we study a class of nonequilibrium lattice models where particles hop in a particular direction, say from a site to its right nearest neighbour, where hop rates not only depend on the occupation of the departure site but also 
on the occupation of all of its neighbours within a range $R$; 
hereafter, we refer to  this process  as the finite range process 
(FRP). We demonstrate that, in one spatial dimension, one can have a CFSS for various specific choices of hop rates;  what we mean by the CFSS  here is that the   steady state   
probability weight can be written as  a product of functions of  
$R+1$ variables, each of them being an occupation number in the 
cluster of $R$ consecutive sites. 
A special case of the CFSS with  $R=1$, called the pair-factorized 
steady state (PFSS), was recently proposed and studied in \cite{PFSS_Evans} where it was shown that, for a particular class of PFSS, 
the system can also undergo  a condensation transition. Later, the PFSS has been found in continuous mass-transfer models \cite{PFSS_Mass, Bertin}, in systems with open boundaries \cite{PFSS_Open} and in random graphs 
\cite{PFSS_Graph}, etc. However, non-trivial spatial structure, which is not present in a FSS, has not been explored before.

We show that, for a broad class of systems having a CFSS with any $R$, there exists a finite dimensional transfer-matrix representation of the steady state. Being finite dimensional, these matrices are quite convenient to manipulate and help in exact calculations of spatial correlation functions of any order. Moreover, we propose a sufficient criterion for the hop rates that can give rise to condensation transition in FRP in general. Surprisingly, we find that a small perturbation to an FSS could destroy condensation transition, if any.

The paper is organized  as follows. In Sec. \ref{sec:model}, 
we discuss the model and its steady state in general for the $(R+1)$-
cluster. In Sec. \ref{sec:2clust}, we formulate a transfer 
matrix method to calculate the correlation functions for $2$-
cluster and then, in Sec. \ref{sec:gen}, we generalize the 
matrix formulation to $R>1$; continuous mass transfer models  are 
also discussed in this section. Some useful applications of the FRP 
in the context of steady state thermodynamics for systems with 
short-ranged correlations, is studied in Sec. \ref{sec:app}.  
Finally, Sec. \ref{sec:conclude} provides conclusions, followed 
by open issues and  discussions.

\section{Model\label{sec:model}}

The model (see Fig. $1$) is defined  on a one dimensional  periodic lattice  with 
sites labeled by  $i=1,2, \dots L.$   Each site  $i$ has  a non-
negative integer variable  $n_i$ representing the number of 
particles at that site (for a vacant site $n_i=0$). Particle from  
any randomly chosen site $i$  can  hop to one  of its nearest  
neighbours, say the right neighbour, with  a rate that depends on   
the number of particles  at  all the sites which are 
within a range $R$ with respect to the departure site:
\begin{eqnarray}
(\dots, n_{i-1}, n_i, n_{i+1},\dots )&&\longrightarrow (\dots, n_{i-1}, n_i-1, n_{i+1}+1,\dots)  \cr
{\rm with} ~{\mathbf{rate }}&&~ u(n_{i-R}, \dots, n_i, \dots, n_{i+R}).~~~
\end{eqnarray}
Clearly the  total number of particles $N= \sum_i n_i$, or the density  $\rho= N/L$,   is conserved by this dynamics. 
\begin{figure}[h]
 \centering \includegraphics[width=8 cm]{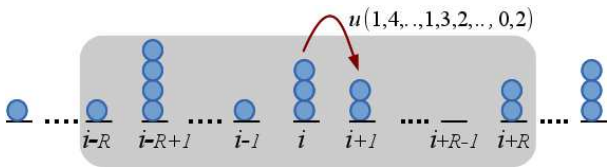}
\caption{(Color online)  In the one dimensional  finite range process 
(FRP)   a particle  hop from a   site $i$ to its right nearest neighbour with a 
rate  that depends on   occupation of   site  $i$  (here  $n_i=3$) and   
all  its neighbours within a range $R$. The  lattice model, for a certain   
hop rate,   can have an  $(R+1)$-cluster-factorized steady state.}
\label{fig:dyn}
\end{figure}
\\
For $R=0,$  this model is identical to the zero range process (ZRP)  
\cite{ZRP_Evans} with hop rate $u(n_i),$  an exactly solvable non-
equilibrium model that evolves to a factorized steady state (FSS)
\be 
 P(\{n_i\}) \propto  \prod_{i=1}^L f(n_i) \delta( \sum_i n_i -N), \label{eq:FSS}
\ee 
with $f(n) = \prod_{m=1}^n u(m)^{-1}.$
The process considered in this paper is a generalized version 
of the ZRP and hereafter we refer to it as finite range process 
(FRP).

For $R>0,$  the steady state of FRP in general cannot have an  
FSS as there are nonzero spatial correlations; however, there can 
be exceptions in specific cases. We  provide explicit  proof, in 
the Appendix, that,  for $R=1,$  the factorized steady state can be 
achieved  only for two  cases -  when  the hop rate  is $u(k,m,n) 
=  v(m)$ or when    $u(k,m,n) =  w(m,n).$   The first case is the 
ZRP and the second one, where  hop rate depends on both   number     
of particles  in  both departure and arrival sites, is known as the 
misanthrope  process  (MP; see \cite{Beyond_ZRP} for a review).  
Since  a FRP with $R>1$ includes  $R=1$ as a special case, one 
expects that, except for the ZRP and the MP, there cannot  be a 
factorized steady state (FSS) for these   classes of systems.

For the FRP, we first try  whether a $(R+1)-$ cluster-factorized 
form, 
\begin{equation}
 P(\{n_i\}) \propto  \prod_{i=1}^L   g(n_i, n_{i+1},\dots n_{i+R}) \delta( \sum_i n_i -N) \label{eq:CFSS}
\end{equation}
with cluster-weight function $g$ of $R+1$ occupation variables,   
can be a steady state weight for Master equation   
\bea
 \frac{d}{dt} P(\{n_i\}) &=&  \sum_{i=1}^L   u(n_{i-R},\dots,n_i,\dots,n_{i+R})  P(\{ n_i\}) \cr
  & - & \sum_{i=1}^Lu(n_{i-R},\dots,n_i+1,n_{i+1}-1,..,n_{i+R}) \cr
  && ~~~~~~~  \times  P(\dots,n_{i-1}+1,n_{i}-1,\dots).
\label{eq:master}
\eea
Now, one can  verify that  a cluster-factorized form  of  steady state, as in  Eq. (\ref{eq:CFSS}),   is indeed possible 
when the hop rate at site $i$ satisfies the following condition  
\begin{eqnarray}
u(n_{i-R}, \dots, n_{i}, \dots,n_{i+R}) ~~~~~~~~~~~~~~~~~~\cr 
 ~~~  ~~~= \prod_{k=0}^R \frac{g(\bar n_{i-R+k},\bar n_{i-R+1+k}, \dots, \bar n_{i+k})}{ g(n_{i-R+k},n_{i-R+1+k}, \dots, n_{i+k})}, \label{eq:Z20}
\label{eq:gen_hop_rate}
 \end{eqnarray}
 where $ \bar n_{j}=n_{j}-\delta_{ji} $. 
A simple way  to   prove this   is to construct a  pair-wise 
balance   $-$ for  every   hop that   takes  configuration $C \to 
C'$  there is  a suitable and unique configuration $C''$ such that 
$P(C) W(C\to C')  = P(C'') W(C''\to C).$ For any configuration   
$C= \{\dots, n_{i-1}, n_i,n_{i+1},..\}$,  a particle hopping  from  
site $i$  can be  balanced  by   taking   $C'' = \{\dots, 
n_{i-1}+1, n_i-1,n_{i+1},..\}$ with  hopping from $i-1.$ Equation 
(\ref{eq:gen_hop_rate}) is    important as it says that  any  
desired   cluster-factorized state  can be  obtained in  FRP  by a  
choosing a suitable  $R$-range  hop rate $u(n_{i-R}, \dots, n_{i}, 
\dots,u_{i+R}).$  

In the  rest of the paper, we  discuss various features  of the  
cluster-factorized steady state and their applications.

\section{2-clusters : Pair factorized steady state (PFSS) \label{sec:2clust}}

Let  us  start with  $R=1$, for which    the steady state is   
factorized as product of $2$-site clusters, commonly known as the 
pair-factorized steady state (PFSS). In this case,  particles  
hop from a site $i$ to $i+1$ with rate  $u(n_{i-1}, n_i,n_{i+1})$      that  depends 
on the  occupation of  departure  site and its neighbours.  To have a 
pair-factorized steady state of the form
\begin{equation}
P(\{n_i\})  = \frac{1}{ Z_{L,N}} \prod_{i=1}^L   g(n_i, n_{i+1})  \delta( \sum_i n_i -N) \label{eq:PFSS}
\end{equation}
with  a canonical partition function 
\bea
Z_{L,N}  =  \sum_{\{n_i\}}   \prod_{i=1}^L   g(n_i, n_{i+1})  \delta \left( \sum_i n_i -N \right), \nonumber
\eea
 the hop rate  must  satisfy  Eq. (\ref{eq:Z20})  with $R=1,$ 
 \begin{equation}
 u(n_{i-1}, n_{i}, n_{i+1}) = \frac{g(n_{i-1}, n_{i}-1)}{g(n_{i-1}, n_{i}) }  \frac{g(n_{i}-1, n_{i+1})}{g(n_{i}, n_{i+1})}. \label{eq:Z21} 
\end{equation}

Unlike the FSS, the PFSS inherently generates spatial correlations 
and, like the FSS, it can lead to real-space condensation for 
certain hop rate \cite{PFSS_Evans}. This  study  has been  later generalized  
on arbitrary graphs \cite{PFSS_Graph}, open boundaries 
\cite{PFSS_Open} and for studying  mass transport processes and 
condensation transition therein for discrete (particle) as well as  
continuous mass \cite{PFSS_Mass}, etc. None  of these studies,  however,
attempted to calculate  the  spatial  correlations in  these  
systems. In fact, the presence of  spatial  correlations  can
change the nature of transitions  by  creating  spatially extended condensates 
with   or without    tunable  shapes  \cite{PFSS_Shape}.

To calculate spatial correlation functions we use  the  transfer 
matrix formulation   which is  possible for a large class of systems having a CFSS. 
For the purpose of illustration we mainly   discuss  this approach   elaborately  for 
the PFSS. Since the PFSS  with  any arbitrary cluster-weight function  $g(n_i,n_{i+1})$ can be  
obtained  from  a suitable hop rate $u(n_{i-1}, n_i,n_{i+1})$  [as in Eq. (\ref{eq:Z21})], 
we rather focus  on the   functional form of $g(n_i,n_{i+1})$, not  
on the hop rate.  In fact, any arbitrary function $g(n_{i},n_{i+1})$  
is  an element 
of the  infinite dimensional  matrix  
\be
G =\sum_{n=0}^\infty  \sum_{n'=0}^\infty g(n,n')  |n\ra  \la n'| \label{eq:G}
\ee
where   $\{|n\ra\}$ are   the  standard  infinite dimensional   basis vectors  
which satisfy a completeness  relation $\la n|n'\ra = \delta_{nn'}$. 
Then, in the  grand canonical ensemble (GCE), where a fugacity $z$ controls the density $\rho$, the   partition sum 
can be written as 
\be
\cZ_L(z) = \sum_{N=0}^\infty Z_{L,N} z^N =Tr[T^L] 
\ee
where  the transfer matrix $T$ has element $\langle n |T|n' \rangle =  z^{(n+n')/2} g(n,n')$. In the thermodynamic limit  $Z_L(z)\simeq  \lambda_{max}^L$ (when  $\lambda_{max},$  the  largest eigenvalue   
of $T$ is non-degenerate). 
Once we know the grand partition sum, we can calculate  various observables; for example, all the moments for occupation number 
$n$ at a site, 
\bea
\langle n^{k} \rangle  = \frac{1}{L}\frac{1}{\cZ_L(z)} \left(z\frac{d}
{dz} \right)^{k} \cZ_L(z).
\eea
For  $k=1$, we get density of the system $\rho= \langle n \rangle 
=\frac{1}{L}\frac{d}{dz} \ln \cZ_L(z);$ by  inverting this density-fugacity relation, one can express  other observables  as  a function of $\rho.$

This  matrix formulation is  quite general  and works for   any 
form of  weight function  $g(n_{i},n_{i+1})$; however managing      infinite dimensional  matrices  is a  challenging task.   
In the  following, we show that,  for a large class of weight functions, one can have a finite dimensional representation which, in some cases, can even be extended to $R>1.$

Let us consider  a weight   function which  has the following form
\be
g(n_i,n_i+1)= \sum_{\kappa=0}^K   a_\kappa (n_i) b_\kappa (n_{i+1})\label{eq:g_gen},
\ee
where  $a_\kappa(n),  b_\kappa(n)$   are arbitrary functions,  not  necessarily analytic. 
It is evident that $g(n_i,n_{i+1})$  can be  
written as  an inner  product of two  $(K+1)$-dimensional vectors,
\bea
g(n_i,n_{i+1}) = \langle\alpha(n_{i})| \beta(n_{i+1})\rangle,
\eea
where
\bea
   \langle \alpha(n)|& =&    ( a_0(n),  a_1(n),\dots, a_K(n))\cr 
  \langle \beta(n)| &=&  ( b_0(n),  b_1(n),\dots, b_K(n)).  \label{eq:ab}
\eea

Then  the  partition  sum   in grand canonical ensemble 
is  $\cZ_L(z) = Tr[T(z)^L] $   with  
\be T (z) = \sum_{n=0}^{\infty}  z^n |\beta(n)\rangle \langle\alpha(n)|
\label{eq:Z}
\ee
a $(K+1)$-dimensional matrix.    Now the partition  sum and   the  stationary 
correlation functions  can be calculated    easily.

To  illustrate this,  let us  consider  a simple  example  by setting 
$K=1,$ $b_0(.) =1 =a_1(.),$ and   renaming functions $a_0(.), b_1(.)$  as $f_0(.), f_1(.)$ respectively.
The    weight function   is now, 
\be
g(n_i,n_{i+1})=  f_0(n_i) + f_1(n_{i+1}) \label{eq:Z1}.
\ee  
which we refer to as  {\it sum-form.}  This   particular choice, {\it  i.e.,} a  pair-factorized steady state with a weight function in sum-form, does  not   lead to condensation transition, which we discuss  later  in Sec.  \ref{sec:IV.B}. Also,  in Sec.   \ref{sec:condensation}, we consider a general case of Eq. (\ref{eq:g_gen}), which  gives  condensation transition, and we   develop  a  possible  criterion for the transition.

For any  functional form of  $f_0(n)$ and  $f_1(n)$ 
we always have  an infinite dimensional representation given by   Eq. (\ref{eq:G}).   However,  interestingly in this case,  we can do away with the infinite dimensional representation and get a simple $2$-dimensional  representation  by taking, 
\bea
 \langle\alpha(n)  |= ( f_0(n),1) ~~{ \rm and }~~ \langle\beta(n)  |= (1,f_1(n)).
\eea
The partition sum in  GCE  is  then ${\cal Z} = Tr[T(z)^L],$ where
\be
T(z)=\sum_{n=0}^\infty z^n
\left( \begin{array}{cc}
f_0(n) & 1 \\
f_0(n) f_1(n) & f_1(n) \\
\end{array}
\right). \label{eq:Z2}
\ee
To   see  how    the   spatial correlation functions  can be obtained,
let us     take a specific  form of  the functions  $f_0(.)$ and $f_1(.),$  
\be
g(n_i,n_{i+1})=\frac{\bar q}{ (n_i+1)^\nu} +  \frac{q}{ (n_{i+1}+1)^\nu}, \label{eq:Z5}
\ee
where parameters  $\nu$  and  $0 \leq q \leq 1$ tune the hop  
rate of particles and $\bar q= 1-q,$  corresponding to   
$f_0(n)/{\bar q} = f_1(n)/q = (n+1)^{-\nu}.$ In this  case, the   
desired hop rate, for which the PFSS with weight-function as in Eq. 
\ref{eq:Z5} is realized, is given by  
\bea
&&u(n_{i-1},n_i,n_{i+1})=\left( 1+\frac{1}{n_i} \right)^{2\nu} \left[ \frac{\bar q n_{i}^\nu +q(n_{i-1}+1)^\nu}{\bar q (n_{i}+1)^\nu +q(n_{i-1}+1)^\nu} \right] \cr 
&&~~~~~~~~~~~~~~~~~~~~~~~~~~~~~~~\times \left[ \frac{\bar q(n_{i+1}+1)^\nu +q n_{i}^\nu}{\bar q (n_{i+1}+1)^\nu +q(n_{i}+1)^\nu}\right].\nonumber
\eea
In the extreme  limits  $q=0$ and  $q=1,$ the model  reduces to zero range process (details will  be discussed in Sec. \ref{sec:app}).

The transfer matrix , following  Eq. (\ref{eq:Z2}),  becomes 
\be
T(z)=\frac{1}{z}\left( \begin{array}{cc}
\bar qLi_{\nu}(z) &q\bar q \frac{z }{1-z} \\
Li_{2\nu}(z) & q Li_{\nu}(z) \\
\end{array}
\right) \label{eq:Z7}
\ee
where $Li_{\nu}(z)$  are  the Polylog functions. The eigenvalues 
of  $T$ are   
\be 
\lambda_\pm=\frac{Li_{\nu}(z)}{2z}\left(1\pm
 \sqrt{(q- \bar q)^2  +\frac{4q\bar q z Li_{2\nu}(z)}{(1-z) Li_{\nu}(z)^2}}\right). \label{eq:Z8}
\ee

The partition function  $ \cZ_L (z) =  \lambda_+^L + \lambda_-^L$      
in the thermodynamic  limit  $(L\rightarrow\infty)$ becomes
$\cZ_L (z)  \simeq   \lambda_+^L$  and thus the density 
\be 
\rho(z)=z \frac{d}{dz} \mathrm{ln} \lambda_+. \label{eq:Z9}
\ee
Throughout the paper, we calculate observables only in the 
thermodynamic limit. Let us  consider  $q=\frac{1}{2}$ and $\nu 
=-1$  (results for  different $q$ and $\nu$ are discussed in 
Sec. \ref{sec:app});  here $\lambda_\pm = \frac{1}{2}{(1\pm \sqrt{1+z})}/{
(1-z)^2}$  and the density  is 
\be 
\rho=\frac{2}{1-z}-\frac{1}{2 \sqrt{1+z}}-\frac{3}{2}.
\ee 
Now  we proceed to calculate the  correlation functions, first the 
two-point correlation function and later the higher order. The  two point 
correlation function is defined by
\bea
C(r)=\langle n_i n_{i+r} \rangle  - \langle n_i  \rangle  \langle n_{i+r} \rangle. \label{eq:corfor}
\eea
For $r>0$ we have
\bea
C(r)&=& \frac{Tr [T'T^{r-1}T'T^{L-r-1}]}{ Tr[T^L]} -\rho^2 \label{eq:2corr}
\eea
where $T'= dT/d(\ln z).$
For  $q=\frac{1}{2}$ and $\nu=-1$, 
we  get 
\be
C(r) = \rho^2  \frac{z(3+z)^2}{4(1+z)(1-z)^2} e^{-r/\xi}
\ee
with  $\xi^{-1}=  |\ln \frac{\lambda_-}{\lambda_+}| = |\ln 
\frac{1-\sqrt{1+z}}{1+\sqrt{1-z}}|$  being the inverse  correlation 
length. The correlation function for $r=0$ is nothing but the 
variance  $\sigma^2(\rho)$ of single-site occupation variable 
$n_i$, i.e.,
\bea
C(0)\equiv \sigma^2 (\rho)= \langle n^2_i  \rangle  - \langle n_i  \rangle^{2}= \frac{Tr [T''T^{L-1}]}{ Tr[T^L]} -\rho^2 
\eea
where  $T''=d^2T/d(\ln z)^2$
and, again for $q=\frac{1}{2}, \nu=-1$,
\be
C(0) = \frac{z}{4(1-z)^2} \left[ \frac{z^2+14z+17}{(1+z)}-\frac{8}{\sqrt{1+z}} \right] .
\ee
Now, we turn our attention to higher order  correlation functions. 
The $3$-point  correlation function, for example, is defined as  
\bea
{\cal C}(r_1,r_2)&=&\langle n_i n_{i+r_1} n_{i+r_1+r_2}
\rangle - \langle n_i \rangle  \langle  n_{i+r_1} \rangle  \langle  n_{i+r_1+r_2} \rangle\nonumber
\eea
which, in terms of transfer matrix, can be evaluated from the 
expression  
\bea
{\cal C}(r_1,r_2)= \frac{Tr [T'T^{r_1-1}  T'T^{r_2-1}  T'T^{L-r-1}]}{ Tr[T^L]} -\rho^3.
\eea
We find that the three-point correlation function can be written in 
terms of the two-point correlation functions as 
\be 
{\cal C} (r_1,r_2)=\rho \left[ C(r_1)+C(r_2) - B(z) C(r_1)C(r_2) \right]
\ee
where $B(z)$ also depend on the parameters $q$ and $\nu$; for  
$q=\frac{1}{2}$ and $\nu=-1$, we get $B(z) =1 +  8z(1+z) /(3+z)^2.$
In a similar  way, one can  calculate  all the higher order 
correlation functions  exactly. 

To conclude, when the  weight function   $g(n_i,n_{i+1})$ is a sum  of  two 
functions as in  Eq.  (\ref{eq:Z1}),  the correlation   length 
$\xi =  |\ln\frac{ \lambda_- }{ \lambda _+}|$  remains finite 
at any density as $\lambda_- < \lambda_+$  for any choice of $q$ and 
$\nu.$

\section{Generalizations\label{sec:gen}}

\subsection{3-clusters and general (R+1)-clusters}

In this section we    consider  some specific   models   of   FRP 
with  $R>1$  which    give rise   to $(R+1)$-cluster-factorized steady state.   Corresponding partition function  in the  
grand canonical ensemble  would  require  contraction of a    
tensor product    which is usually a  hard task \cite{Tensor}. 
Our  aim  here  would be   to  obtain, if  possible,  a  matrix formulation  that  can  accommodate    some  cluster-
factorized   steady states for  any  $R>1$.   For $R=2$ we have a  $3$-site  cluster factorized steady state, 
\bea
P(\{ n_i\}) = \prod_{i=1}^L  g(n_i, n_{i+1}, n_{i+2})  \delta\left(\sum_{i=1}^L n_i -N\right).\nonumber
\eea
For illustration we consider  a cluster-weight function,
\be 
g(n_{i},n_{i+1},\dots n_{i+R})=\sum_{\kappa =0}^R f_\kappa(n_{i + \kappa}) \label{eq:Z3}
\ee
which  is a  simple generalization of  the sum-form  given in  Eq.  (\ref{eq:Z1}).    
We will    now show that   a  grand  partition function   of  a 
finite range process     which has a     $(R+1)$-cluster-factorized steady  state    with  a  
weight function   given by Eq. (\ref{eq:Z3}) can    be written  
as $\cZ_L(z) =Tr[T^L]$ where    $z$ is the fugacity  and $T$ is a 
$2^{R}$-dimensional   transfer matrix.  Since we intend to  obtain 
the transfer matrix  for iteratively, let us rewrite    the  
transfer  matrix  given  by Eq.   (\ref{eq:Z2})   for $R=1$   in a 
convenient form,  
\be
T_1(z)=\sum_{n=0}^\infty z^n {\bf F}_1(n); ~
{\bf F}_1(n)=
\left( \begin{array}{cc}
f_0(n) & 1 \\
f_0(n) f_1(n) & f_1(n) \\
\end{array}
\right) \label{eq:F1}
\ee

In a similar way, we  extend  to  $R>1$   and  write 
$T_{R}=\sum_{n=0}^\infty z^n  {\bf F}_R(n)$ where  the $2^{R}$-
dimensional matrix can be written as 
\be 
{\bf F}_{R}= \left( \begin{array}{cc}
{\bf F}_{R-1} & A_{R-1}{\bf F}_{R-1} \\
f_R {\bf F}_{R-1} & f_R A_{R-1} {\bf F}_{R-1} \\
\end{array}\right), \label{eq:FR}
\ee
using a constant matrix    
$$A_{R}=\left( \begin{array}{cc}
0 & 0 \\
I_{2^{R-1}} & 0 \\
\end{array}
\right),
$$ \\
where  $I_{2^{R-1}}$ is a  $2^{R-1}$-dimensional identity matrix.
For $R=0$,  we take  $A_0 =1.$ Since $R=0$ corresponds to the ZRP  
which has a    factorized steady state, we have  ${\bf F}_0(n)=f_0(n),$ 
which  is a    scalar. Clearly ${\bf F}_1$  in Eq. (\ref{eq:F1})    
satisfy  Eq. (\ref{eq:FR}).   A little  more algebra would show 
that     the transfer matrix for $R=2$ is     
\be 
T_2=\sum_{n} z^n \left( \begin{array}{cc}
{\bf F}_1 & A_1 {\bf F}_1 \\
f_2{\bf  F}_1 & f_2 A_1 {\bf F}_1 \\
\end{array}
\right)=\sum_{n} z^n {\bf F}_2(n). \label{eq:Z4}
\ee
From  the  transfer matrix, one  can,  in principle, calculate the 
expectation   value  of  any desired observable.  We  will not 
discuss further the  finite range  process $R>1$; the finite 
dimensional  transfer matrix  is  expected to   generate   spatial  
correlations which was absent in the ZRP.   We discuss some of the 
models in details  which undergo condensation transitions
(see Sec. \ref{sec:app}).

\subsection{ Continuous mass model }

Until  now, we have studied CFSS  on a   one dimensional lattice  
with  each site  having  a discrete  variable,  called  the 
occupation variables or number of  particles. The model and the 
matrix formulation can be extended,  without any particular 
difficulty, to  systems with  continuous mass $m$. As an example, 
let us consider  
\be 
g(m_{i},m_{i+1},m_{i+2})=m_i+m_{i+1}+m_{i+2}. 
\ee
A $3$-cluster-factorized steady state with the above weight-function  
can be obtained    when  $\epsilon$ amount of mass is transferred  
 from site $i$ to  $i+1$ with the  rate 
 \bea
&& u(m_{i-2},m_{i-1},m_i,m_{i+1},m_{i+2}) \cr
 &&~~~~= \prod_{k=0}^2 \left[1- g(m_{i-2+k},m_{i-1+k},m_{i+k} )^{-1} \right].
 \eea
For small $\epsilon$,   the     model is equivalent   to a discrete 
model where mass is measured in units of $\epsilon.$   In fact, the 
residual mass (actual mass modulo $\epsilon$)  at   any site   does   not 
change  during evolution. The residual masses,   each  being   smaller than   a pre-defined 
value $\epsilon$ which can be  made arbitrary small, 
does not  contribute  to  the asymptotic form of the hop rate. 
Thus  we   would obtain a transfer matrix   $T_2$   discussed in the 
previous section,  with    $f_{0,1,2}(m) = m,$  but the sum 
$\sum_m$  will now be replaced by an integral $\int dm.$
Defining, a  chemical potential $\mu$  (where $z=e^{\mu}$), we   
get the transfer matrix, as in Eq. (\ref{eq:Z4}), 
\be 
T(\mu)=\frac{1}{\mu^2} \left( \begin{array}{cccc}
1 & \mu & 0 & 0 \\
\frac{2}{\mu} & 1 & 1 & \mu \\
\frac{2}{\mu} & 1 & 0 & 0 \\
\frac{6}{\mu^2} & \frac{2}{\mu} & \frac{2}{\mu} & 1 \\
\end{array}
\right).
\ee
This matrix has eigenvalues  $\frac{1}{\mu^2}\{ \lambda, \lambda_1 
e^{\pm i \theta}, \lambda_2\},$ where $\lambda$ (the  
largest eigenvalue), $\lambda_1$, $\theta$ and $\lambda_2$ are 
independent of $\mu$, and their approximate numerical values are 
$\lambda \approx 3.86841$,    $\lambda_1 \approx 1.10465,$  $\theta 
\approx 1.87254$ and  $\lambda_2 \approx -0.21184.$ In the 
thermodynamic limit, the partition function is $\cZ_L = \left( 
\lambda/\mu^2 \right)^L,$ and  density $\rho=-{2}/{\mu}.$ The 
two-point correlation function for $r>0$ is
\bea
C(r)&=&\langle m_i m_{i+r} \rangle -\rho^2  \cr
 &=&\rho^2\left[c_{2} \left(\frac{\lambda_2}{\lambda}\right)^{r}+2 c_{1}\left(\frac{\lambda_1}{\lambda}\right)^{r} \cos(r \theta+\alpha) \right] 
\eea
 where,  $c_1=0.3380 $,  $c_2= -0.0375$  and $\alpha=0.1804.$ And, for $r=0$, the correlation (actually  $\sigma^2$) is
 $C(0)=\langle m^2 \rangle -\rho^2=   0.6704 \rho^2.$

\section{Applications \label{sec:app}}

\subsection{Condensation transition \label{sec:condensation}} 

One important  feature in these simple one dimensional models is 
that they can exhibit condensation transition at a finite 
density when one or more parameters  in  the rate functions are 
tuned.  To  demonstrate the possibility of  a   condensation 
transition in the CFSS, for any $R$, we consider the weight of 
$(R+1)$-cluster to  be, 
\be
g(n_i,n_{i+1},...,n_{i+R})= \frac{\left[q+ \sum_{j=0}^{R}  n_{i+j} \right]^K}{(n_i+1)^\nu}, \label{eq:g_condensation}
\ee
where   $K, \nu$ and $q$ are  positive and $K$ is an integer.  
This  steady state 
weight can be generated from a hop rate given by Eq. 
(\ref{eq:gen_hop_rate}),
\be
u=\left( 1+\frac{1}{n_i} \right)^\nu \left[ \prod_{k=0}^R \left(1-\frac{1}{q+\sum_{j=0}^{R} n_{i-j+k}} \right) \right]^K .  \label{eq:uRcondensation} 
\ee

{\it Case with $R= 1$ (PFSS). $-$} We  first   consider  $R=1.$  It is easy to see that  for any $K,$     the weight function  Eq.  (\ref{eq:g_condensation})   
can be expressed     as    Eq. (\ref{eq:g_gen})   with suitable  choice  of $a_\kappa(n)$ and  $b_\kappa(n)$  where   $\kappa$  varies from $0$ to $K,$ leading to a $(K+1)$  dimensional transition matrix.  We further set the parameters $K=1=q;$ this  gives rise  to  a PFSS, as  in Eq. (\ref{eq:PFSS}), with $g(n_i,,n_{i+1}) = (n_i+n_{i+1}+1)/(n_{i}+1)^\nu$, which can be realized when a particle hops out from a site $i$ (to the right neighbour), having $n_i>0$ particles, with the following rate 
\be
u=\left(1+\frac{1}{n_i}\right)^\nu  \frac{n_i +n_{i-1}}{ 1+n_i +n_{i-1} }    \frac{n_i +n_{i+1}}{ 1+n_i +n_{i+1} } \label{eq:Z101}.
\ee 
For this case, we    can   obtain  exact results  following the   
matrix formulation developed   here. First we  write  $g(m,n)= 
\langle \alpha(m)| \beta(n) \rangle$  where $ \langle \alpha(m) |
=\left(  (m+1)^{-\nu}, (m+1)^{1-\nu}  \right)$,
$ \langle\beta(n) |=\left(n,1 \right)$. Thus the grand 
partition function can be written as $Z(z)= Tr(T^L)$ with 
\bea
T  = \sum_{n=0}^\infty  |\beta(n) \rangle \langle \alpha(n)| z^n =  \frac{1}{z}
\left( \begin{array}{cc}
Li_{\nu-1}(z) & Li_{\nu-2}(z)  \\
  Li_{\nu}(z)& Li_{\nu-1}(z)
\end{array}
\right)
\nonumber
\eea
The   eigenvalues of  $T$ are  
\bea
\lambda_\pm(z)  =  \frac{1}{z} \left( Li_{\nu-1}(z) \pm\sqrt{ Li_{\nu}(z) Li_{\nu-2}(z)}\right), \nonumber
\eea
which   leads  to the density-fugacity relation     $\rho(z)= z \lambda_+'(z)/\lambda_+(z)$   and 
the   critical    density     $\rho_c =\lim\limits_{z\to1} \rho(z).$ 
It turns out that  for  $\nu\le 4,$  $\rho_c$  diverges  
$-$ indicating  a fluid phase for any density.  For $\nu>4$   we      get, 
\bea
\rho_c&=& \frac{\xi_1(\nu-1) - 2\xi_2(\nu)+\xi_3(\nu)}{2 \xi_2(\nu)+ 2 \zeta(\nu-1)\sqrt{\xi_2(\nu)}  }
+\frac{\zeta(\nu-2)- \zeta(\nu-1)   }{ \sqrt{\xi_2(\nu)}+ \zeta(\nu-1)  }\nonumber
\eea
where   $\xi_k(\nu) = \zeta(\nu) \zeta(\nu-k)$  and  $\zeta(\nu)$ 
are Riemann zeta functions.   Thus, for   $\nu>4$ we have a 
condensate  when density   exceeds this critical value.  Unlike   
the ZRP,  where particles at different sites are not correlated, 
here  we have  non-vanishing  correlation  that  extends   up to a   length  scale 
$\xi(z)= |\ln\frac{ \lambda_-(z)}{\lambda_+(z)}|^{-1}$  which is finite throughout.

{\it Case with $R\ge 2$ (CFSS). $-$}
It is straightforward to extend the matrix formalism to $R>1$   when 
$K=1.$  First, let us  take  $\nu =0.$  In this case,  the  weight function  $g$  takes  a sum-form 
given by  Eq. (\ref{eq:Z3}), for which we  have already  constructed a general transfer-matrix. For $\nu>0,$ the dimension of the transfer matrix remains  the 
same as in  $\nu=0;$  it is only that  each   element  of ${\bf F}_R$ in 
Eq.  (\ref{eq:Z4}) will  be multiplied  by an extra factor 
$(n_i+1)^{-\nu}.$    We omit   the  exact analytic expressions of the  density-fugacity  relation   and the   critical density - the calculations 
are straightforward   but   the  expressions are  very long.   Only  the   numerical  
values  of  critical density  are  tabulated  in  Table \ref{tbl} for different parameters.

\begin{table}
\caption{ Critical density $\rho_c$ for $K=1.$  }
\begin{tabular}{cccc} \hline \hline 
$~~~~~~$& $~~~q=1~~~$ & $~~~q=1~~~$ &    $~~~q=2~~~$ \cr
 $~~~\nu~~~$& $~~~R=1~~~$ & $~~~R=2~~~$ &    $~~~R=1~~~$ \cr
 \hline 
 \hline  
 5 & 0.3254 & $ \infty $ &  0.1591   \cr 
 6 & 0.1054 &0.2773 & 0.0544  \cr
 7 & 0.0429 &0.0981 & 0.0228  \cr 
 \hline
\end{tabular}
\label{tbl}
\end{table}

{\it Criterion  for condensation transition.$-$}
For the ZRP,  it is well known that, provided  the hop rate $u_0(n)$ has an asymptotic form   
\be
u_0(n) =1+  \frac{b}{ n^{\sigma}}  + \dots \label{eq:criteria}
\ee
condensation occurs at a finite density,  when    $\sigma <1,$  or
when $\sigma =1$ but $b>2.$ This criterion  can be  extended to   
any other system (without any constraint on occupation number) when 
the steady state has a factorized  form  (\ref{eq:FSS}); one needs 
to consider and effective  rate function $u_0(n) \equiv 
f(n-1)/f(n)$  and find  its  asymptotic form.  This   criterion  
determines  whether a model can  undergo a  condensation    
transition   and  helps   in understanding phase coexistence   
in hardcore lattice gas models \cite{Criteria, Bijoy}.

Such a criterion for cluster-factorized steady state would be very 
useful.  At present, we  do not have a  general criterion, but  
the examples studied  above    suggest   a sufficient condition  
for  CFSS  to  have    condensation. If the  
rate function can be  expanded as 
$$
u(n_{i-R}, .., n_{i+R}) = \sum_{\nu=0}^{\infty} \frac{B_\nu (n_{i-R}, .., n_{i-1},n_{i+1} .., n_{i+R})}{  n_i^\nu},
$$ 
the condensation  transition can occur when both the conditions  
\bea
(i) ~&&  {\rm both }  \:   B_0  \:{\rm and }  \:B_1   \: {\rm  are ~  constant} \cr
 ~(ii)&& B_1/B_0 >2 \label{eq:pfss_criteria}
\eea 
are satisfied. This  is   only a simple generalization of the    
criterion of condensation in the ZRP. Effectively, $B_1/B_0$ plays  
the role of $b$  in  Eq. (\ref{eq:criteria}). As the hop 
rate in Eq. (\ref{eq:uRcondensation})  can be expanded as 
\bea
u(\dots,n_{i-1}n_i,n_{i+1}\dots ) =  1+  \frac{\nu - K(R+1)}{n_i}   + {\cal O} (\frac{1}{n_i^2}),  \nonumber
\eea
and thus $B_0=1$ and  $B_2= \nu - K(R+1),$ the criterion   
correctly  predicts the condensation which occurs only when 
$\nu>K(R+1)+2.$ This  is same as the usual  condensation 
criterion in the ZRP   if   we treat  $b\equiv B_2/B_0.$ 
In this particular case, we have also checked that   moments 
$\langle  n^k \rangle$ as a  function of $z$,  in leading order, 
are  the same as that in the ZRP  with corresponding  $b$  (see Eq.  
(\ref{eq:criteria}) ). This  criterion, however, cannot be 
applied   to  some of the  following cases studied recently, such 
as, the misanthrope   process  \cite{Beyond_ZRP} and the PFSS     
\cite{PFSS_Evans}. For the first case, $B_0$ and $B_1$ are not 
constants and, for the later case, hop rates are not  analytic 
functions. A criterion of condensation, which  can 
apply to  a cluster-factorized steady state in general is  
desirable and  remains a challenge.

\begin{figure}[h]
\centering \includegraphics[width=8.2 cm]{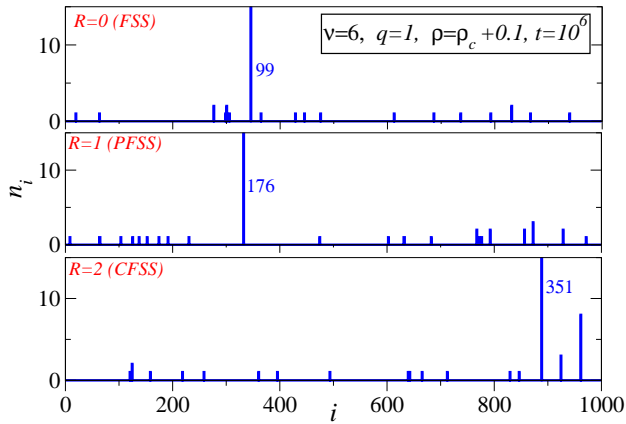}
\caption{  (Color online)  Particle distribution   in   FRP  with  weight function  (\ref{eq:g_condensation}) 
after   $t=10^6$  MCS, starting from a random distribution of particles. Density is  $\rho=\rho_c+0.01.$
The  critical density for  $R=0$  is  $\rho_c=0.01925$; the same for  $R>0$ are  taken from 
table \ref{tbl}.  Clearly, for all cases, the condensate  is  localized to a  single site. 
The condensate size is  written  beside the condensate site.  
}
\label{fig:single_site}  
\end{figure}

We  end  this section with the following remark.   The   
condensation transition   here is      different from   that 
obtained for PFFS  by  Evans {\it et. al.} \cite{PFSS_Evans}.
There, one  observes an extended condensate where   both the size   and the  spatial extent   of condensate scales  with  system size as $\sqrt L.$ This  indicates that  the transition  is associated with a  diverging 
spatial correlation length. Whereas  for the PFSS  (and the CFSS)  
studied here,  the   correlation length remains finite throughout and the transition is 
characterized  by a   diverging   mass fluctuation, as in the ZRP. 
The condensate  is  also localized     to a single site
(see Fig.  \ref{fig:single_site}).  A  detailed  comparison of  
nature of condensate  would   be reported elsewhere \cite{ComingSoon}.

\subsection{ Pair factorized  state with weight function in sum-form\label{sec:IV.B}}

In this section, we first show that  a  pair factorized  steady state with   weight function   $g(n_i,  n_{i+1}) = f_0(n_i) + f_1(n_{i+1})$, 
which we refer to as   {\it sum-form,}  cannot give rise to 
condensation. Then, we demonstrate this considering a
perturbation to a ZRP that converts the existing 
factorized steady state of the ZRP to a PFSS  with  weight function in  the sum-form. For the PFSS with weight function  in  the sum-form, the transfer matrix $T(z)$ is given by Eq. (\ref{eq:Z2}).

The largest   eigenvalue  of  the   matrix  
$\lambda_+ =   \frac{1}{2} (  T_{11} + T_{22}+  \sqrt {   (T_{11}+ T_{22})^2 - 4 {\cal D}},$ where 
${\cal D}$ is the determinant of   $T$  can be   used in Eq.  (\ref{eq:Z9})  to get the density  
$\rho(z).$ With some straightforward algebraic manipulations, one can show  that the  maximum  density at 
$z=z_c=1$ is, 
\bea
\rho_c =\lim_{z\to 1}   \rho(z) = \lim_{z\to 1}\frac{1}{2} \left[   \frac{1}{T_{21}}   \frac{d T_{21}}{dz~}    + \frac{1}{1-z}\right].\nonumber
\eea
Clearly  $\rho_c$   diverges  independent of  the first  term, leading  to a conclusion that   there can not be a   condensation transition at any finite density. Thus, a PFSS {\it cannot} have condensation transition if the  weight function has a sum-form. 
To illustrate this,  we  consider a   simple   zero range  process 
with weight function  $f(n) =  1/(n+1)^\nu$  or   hop rate  $u(n)= f(n-1)/f(n)  =(n+1)^\nu/ n^\nu $
and add a  perturbative term   get a   new  weight function 
\be
g(n_i, n_{i+1}) = (1-q)   f(n_i) +  q f(n_{i+1})
\ee
which depends on occupation  of two consecutive sites. Here   $ 0\le q\le 1 , \bar q =  1-q $ and we  choose $f(n) =  1/(n+1)^\nu.$ 
A  pair-factorized state, as in 
Eq. (\ref{eq:PFSS}), with the above weight function occurs
when particles hop rate is 
\bea
u(n_{i-1},n_i,n_{i+1}) =\frac{  \bar qf(n_{i-1}) +  q f(n_{i}-1)}{\bar q  f(n_{i-1}) +  q f(n_{i})} \cr 
\times \frac{ \bar q  f(n_i-1) +  q f(n_{i+1})}{ \bar qf(n_i) +  q f(n_{i+1})}. \nonumber
\eea

For   both $q=0$ and   $q=1$   we   have a factorized steady state, as in Eq. (\ref{eq:FSS}), which corresponds  to 
 the  ZRP   with particle hop  rate 
\be
u(n) = \frac{f(n-1)}{f(n)}= \left( 1+\frac{1}{n} \right) ^\nu \simeq  1+\frac{\nu}{n}   + {\cal O} (\frac{1}{n^2}).
\ee
Thus    we expect a  condensation transition    for $q=0,1$ when   
$\nu>2$   and the   density    is larger than  a
critical  value $\rho_c.$   In this case the  $\cZ(z) =  F(z) ^L$   
(the transfer matrix  $T(z)$   reduces to a scalar), where 
$F(z) = \sum_{n=0}^\infty h(n) z^n = Li_\nu(z).$ The density is  
$\rho =   z \frac {d}{dz} F(z)$ and thus the critical density  for 
$q=0,1$ is 
\be
\rho_c = \lim_{z\to 1} \rho(z)=\left\{ \begin{array}{ll}
\infty  & \mathrm{for} \: \nu\leq 2, \\
\frac{\zeta(\nu-1)}{\zeta(\nu)}-1 & \mathrm{for} \: \nu >2. \\
\end{array}\right.
\ee

\begin{figure}[h]
\centering \includegraphics[width=8.2 cm]{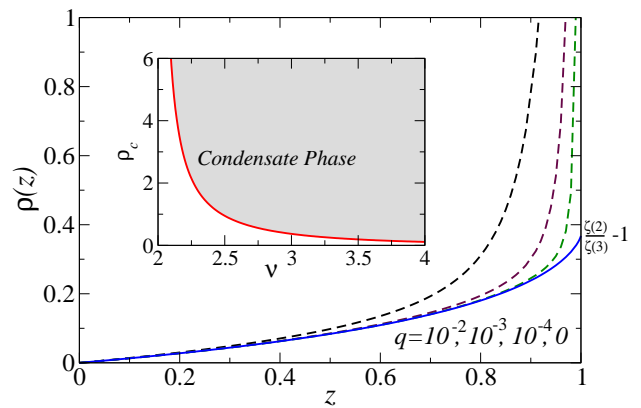}
\caption{  (Color online)  Small perturbation  to the ZRP:  
For  small   $q= 10^{-2},10^{-3}$ or $10^{-4}$,  density $\rho(z)$  
diverges  when $z\to1.$   However for  $q=0$  or for $q=1,$   
$\rho_c = \rho(1) =\frac{\zeta(\nu-1)}{\zeta(\nu)}-1$ is finite, 
leading to a condensation transition   when $\rho> \rho_c.$ Inset    
shows the phase diagram for $\nu=3.$  }
\label{fig:rho_c}
\end{figure}

The  phase-diagram of the  condensation transition  in the $\rho$-$\nu$ 
plane  is shown  in Fig. \ref{fig:rho_c}.  The critical line 
$\rho_c$   separates the condensate phase  from the fluid phase. 
For a general $0<q<1$, we need to calculate the density  using Eqs. 
(\ref{eq:Z7}), (\ref{eq:Z8}), and (\ref{eq:Z9}),
\bea
\rho(z) =\frac{1}{a(a-(1-z)Li_{\nu}(z))} [Li_{\nu-1}(z)(\bar q^2(1-z)^2Li_{\nu}(z)-\cr (1-z)a)+2\bar qqz(Li_{2\nu}(z) +(1-z)Li_{2\nu-1}(z))]-1  \cr 
{\rm  where }~~~~~~~~~~~~~~~~~~~~~~~~~~~~~~~~~~~~~~~~~~~~~~~~~~~~~~~~~~~~~~~~~\cr
a(q,z)=\sqrt{(q-\bar q)^2(1-z)^2Li_{\nu}^{2}(z)+4q\bar q z(1-z)Li_{2\nu}(z)} . \nn
\eea
In the limit  $z\to 1,$ $\rho(z)$ diverges for all  $\nu>0$ and 
thus  the condensation transition is destroyed.
It is somewhat surprising why  for any non-zero $q$ however small, the condensation transition is destroyed. 
It seems  that this  perturbation, which takes the factorized steady state of the ZRP to a pair-factorized steady 
state, forces the condensation to disappear. One could understand 
this  following  the criterion (\ref{eq:pfss_criteria}). For $\nu=3$, the rate for general $q$ has an asymptotic form ({\it i.e.} when  $n_{i} \to \infty$)
)\bea
u(n_{i-1},n_i,n_{i+1})=1+3 \frac{q^2(1+n_{i-1})^3+\bar q^2 (1+n_{i+1})^3}{q  \bar q n_{i}^4 } . \nn
\eea
Thus, here $B_1/B_0=0$ and therefore we should not expect 
condensation for this hop rate.  It can be shown easily that for any   $\nu \geq 1$  
 the asymptotic form of the hop rate  does not satisfy  condition $(ii)$ of ansatz  (\ref{eq:pfss_criteria})  ruling out the
possibility   of   a   condensation transition.

\subsection{Subsystem mass distribution}

It was argued in   recent works  \cite{Bertin, gamma, zero-th} that,  
for systems  with short-ranged interaction, irrespective  of whether they 
are in equilibrium or not,  one could obtain  a state   function 
which plays  the   role of a free energy  function.  It was  shown 
in \cite{gamma, zero-th}
that the steady state distribution $P_v(m)$ of mass in  
subsystems  of volume $v \gg \xi$  can be determined from  the 
functional dependence of the scaled variance  $\sigma^2 (\rho) = (\langle m^2 \rangle - \langle m \rangle^2)/v$, in the limit of large $v$, on  the mass density $\rho$. When $\sigma^2(\rho) \propto \rho^2$ is a
quadratic function of density $\rho$,  the  subsystem mass  
distribution   can be   characterized through a gamma distribution,  
i.e., $P_v(m) \propto  m^{\eta-1}  \exp (\mu m)$, where $\mu = - 
\eta/ \rho $  is an equilibrium-like chemical potential. The 
exponent $\eta$ however depends on the  details  of the model and    
it can be calculated from  the knowledge of  two point-correlation 
function only. The matrix formulation  developed here for the CFSS  
can thus help in determination of $\eta.$

To illustrate this, let us consider a continuous finite range 
process  with $R=1,$ and calculate explicitly the  variance of   
the  subsystem  mass. Consider the following homogeneous 
weight function for a pair-factorized steady state, 
\be
g(m_i,m_{i+1}) = m_{i}^\delta + c~ m_{i}^\gamma m_{i+1}^{\delta-\gamma} 
\ee
The grand partition sum is ${\cal Z}= Tr[T^L]$ where the transfer  matrix $T(\mu)$ ($\mu=\ln(z)$ is the corresponding chemical potential) is given below
\begin{equation}
T(\mu)=\frac{1}{\mu^{1+\delta}} \left( \begin{array}{cc}
\Gamma(\delta+1) & c \frac{\Gamma(\gamma+1)}{\mu^{\gamma-\delta}} \\
\frac{\Gamma(2 \delta-\gamma+1)}{\mu^{ \delta-\gamma}} & c \Gamma(\delta+1) \\
\end{array}
\right) ,
\end{equation}
where   $\Gamma(.)$  are   Gamma functions. 
Eigenvalues  of  $T(\mu)$  are  $\lambda_{\pm}=\Lambda_{\pm}(\delta,\gamma,c)/ \mu^{1+\delta}$ where
\bea
&&2 \Lambda_{\pm}(\delta,\gamma,c)= (1+c) \Gamma(\delta+1)\cr
&& ~~~~~~~~\pm 
\sqrt{(\delta+1)^{2}(1-c)^{2}+4c\Gamma(2 \delta-\gamma+1)\Gamma(\gamma+1)} \nn
\eea
and the  particle  density is 
\be 
\rho=\frac{\partial}{\partial 
\mu}\mathrm{ln}\lambda_{+}=-\frac{\delta+1}{\mu}, \label{rho-mu}
\ee 
implying a fluctuation-response (FR) relation
\be 
\frac{d \rho}{d \mu} = \sigma^2(\rho), \label{FR}
\ee  
analogous to the fluctuation-dissipation theorem in equilibrium.
Now, as shown below, one can check the above FR relation by explicitly calculating both sides of Eq. (\ref{FR}). The r.h.s of Eq. (\ref{FR}) can be calculated by integrating two-point correlation function $\sigma^2(\rho)=\sum_{r=-\infty}^{r=\infty} C(r)$, using Eq. (\ref{eq:2corr}), 
\be
C(r)= \langle n_i  n_{i+r}\rangle  -\rho^2=  \rho^2\left[  A(r)-1 \right]
\ee
where, for   $r>0,$
\bea
A(r)= 1+\left( \frac{\Lambda_-}{\Lambda_+} \right)^r \frac{  (\delta-\gamma)^2/(\delta-1)^2 }{1-\frac{\Gamma(\delta+1)^2}{\Gamma(2\delta-\gamma+1)\Gamma(\gamma+1)}}
\nonumber 
\eea
and 
\bea 
A(0)= \frac{\delta +2}{\delta +1}+\frac{2c}{\Lambda_+}  \frac{(\delta-\gamma)^2}{(\delta +1)^2}
\frac{\Gamma(2\delta-\gamma+1)\Gamma(\gamma+1)}{2\Lambda_+-(1+c)\Gamma(\delta +1)}. \nn
\eea
In this system,  the {\it  gap} $(\lambda_+ - \lambda_-)$ between 
the two eigenvalues is nonzero  and  the  correlation length $\xi=|
\ln \frac{\Lambda_-}{\Lambda_+}|^{-1}$ is finite.  
Therefore, following Ref. \cite{gamma}, the  subsystem  mass distribution $P_v(m)$, for $v \gg \xi$, is a gamma distribution 
where the exponent $\eta$ can be written, using   Eq.  (\ref{eq:Z9}), as
\bea
\eta^{-1}&=&\sum_{r=-\infty}^{\infty} (A(r)-1), 
= \frac{1}{\delta+1}.
\eea
Note that  the  exponent  $\eta$ depends only on the homogeneity exponent $\delta$ but neither on $\gamma$ nor  on $c.$ 
The left-hand side, the compressibility $d\rho/d\mu$, of Eq. (\ref{FR}) gives the same $\eta= \rho^2 (\frac{d\rho}{d\mu})^{-1} = \delta+1$, by differentiating the expression $\rho= -(\delta+1)/\mu$ in Eq. (\ref{rho-mu})  with respect to $\mu$; this is a proof that the fluctuation-response relation indeed holds here and also is consistent with the additivity  property proposed earlier for these systems \cite{Bertin, gamma}.

In  principle, the single-site mass distribution (for $v=1$) can    
be calculated  straightforwardly  from  the  moments, but the exact   
closed form expression is  hard to obtain. In this regard,  this 
formulation  \cite{gamma, Arghya} for obtaining the subsystem mass 
distribution from the two-point correlation function is  quite  useful  
in obtaining  the macroscopic  behaviour  of the system.

\section{Summary \label{sec:conclude}}

We have introduced a class of nonequilibrium finite range processes (FRP) where particles on a one dimensional periodic lattice can hop in a particular direction, from one site to one of its nearest neighbours, with a rate that depends on the occupation  of all the  sites within a range  $R$ starting from the 
departure site.  We show that,  for certain  specific functional forms 
of the hop rates, the FRP has a cluster-factorized steady state (CFSS), i.e.,  the steady state probability of a microstate can be written as a  product  
of  cluster-weight functions $g$   having  $(R+1)$  arguments ~-
~the occupation  numbers of  $(R+1)$  consecutive  sites. The model with $R=0$ reduces to the familiar zero range process (ZRP), which has factorized steady state.

The CFSS with $R=1$ reduces to the pair-factorized steady state (PFSS) and its steady state can always be represented by an infinite-dimensional transfer matrix. However, for the CFSS with $R>1$, a matrix formulation is {\it not} guaranteed.
In this work, we show that, for a large class of systems having CFSS with $R>0$, there exists a {\it finite} dimensional matrix representation. Being finite dimensional, these matrices are easy to manipulate and thus help in exactly calculating the $n$-point correlation functions for any $n$. The two-point correlation function ($n=2$) can be utilized to characterize the subsystem mass distribution in these nonequilibrium systems in terms of a nonequilibrium chemical potential and a free-energy function, which are obtained through a fluctuation-response relation \cite{Bertin, gamma} - analogous to the equilibrium fluctuation-dissipation theorem.

Even though the transfer-matrix is finite dimensional, the CFSS can undergo a condensation phase transition. We obtain a sufficient  condition  for the condensation  transition for a particular class  of hop rates in the FRP in general. The nature of the condensation transition studied in this paper are however different from those studied in systems having a PFSS \cite{PFSS_Evans}; the condensate here remains localized, as in the ZRP, in contrast to the extended condensate observed in  \cite{PFSS_Evans, PFSS_Mass, PFSS_Graph}. Moreover, the condensation transition studied here occurs solely due to the diverging  mass fluctuations at certain critical density $\rho_c,$  not due to a diverging correlation length; in fact,  the correlation length remains finite throughout.

We should  mention that   it is {\it  always}  possible to 
construct a hopping dynamics of FRP so that  it evolves to a 
desired steady  state which  is cluster-factorized. However,
for a given hop rate, there is no simple way {\it to check} if 
it can give rise to a CFSS.  In our opinion, the FRP is a very general 
class of models as it  includes the Ising model, Potts model,    
misanthrope process, urn models,  symmetric and asymmetric   
exclusion processes on a ring and many other models 
(one of them, of course, zero range process). More  
importantly, the method developed here,  could help in    
finding the exact  steady state structure  in models   even  when  
the interactions extend  beyond two sites.

%

\section*{Appendix}

In this  Appendix,  we provide an  argument that FRP can   have a  factorized steady state   only for  $R=0$ (namely the ZRP) and 
 for  some specific misanthrope process  (special cases of $R=1$).    For any $R>1,$    however,  one cannot have a  factorized  
 steady state  in general.
First  we   consider  $R=1$  and show that, in this case,  the   hop rate reduces  to those in the ZRP or in the misanthrope process, when   one demands  a    factorized steady state.   One   can  construct a general   proof   in a similar way,    
that   condition of FSS     would  reduce   the  hop rate  of 
 FRP with  $R>1$ to the ZRP or the misanthrope process.  In  Appendix B, we provide a proof  of the above  
for  the  hop  rates which  can be written in a product form.  

\section*{Appendix A : NO FSS FOR $R=1$}

In this section, we show that, for $R=1$, one cannot have 
a factorized steady state for the general hop rate 
$u(n_{i-1},n_i,n_{i+1})$. 
The Master equation for FRP   for  general $R>0$ is  
\bea
\frac{d}{dt} P(\{n_i\}) =  \sum_{i=1}^L F(n_{i-R},\dots,n_{i},\dots,n_{i+R}), \label{eq:master}
\eea
where
\bea
&&F(n_{i-R},..,n_{i},..,n_{i+R})= u(n_{i-R},..,n_i,..,n_{i+R})  P(\{ n_i\})\cr
 &&~~~~~~~~~~~~~~~~-u(n_{i-R},..,n_i+1,n_{i+1}-1,..,n_{i+R}) \cr
 &&~~~~~~~~~~~~~~~~~~~~~~~\times  P(..,n_i+1,n_{i+1}-1,..). \label{eq:F}
\eea
In the steady state,  right  hand side  of Eq. (\ref{eq:master})  must vanish, which  can happen  if
\bea
&&F(n_{i-R},\dots,n_{i},\dots,n_{i+R})=h(n_{i-R},\dots,n_i,\dots,n_{i+R-1})\cr
&&~~~~~~~~~~~~-h(n_{i-R+1},\dots,n_{i},\dots,n_{i+R}) \label{eq:h}
\eea
for  some   arbitrary    function $h$ of $2R$ arguments.  Note, that  the above  cancellation scheme  
is only a  sufficient condition.

Now let us consider  $R=1,$  and demand that the steady state  has a factorized form   given by Eq. (\ref{eq:FSS}). Then 
\bea
u(n_{i-1},n_i+1,n_{i+1}-1)\frac{f(n_i+1)}{f(n_i)}\frac{f(n_{i+1}-1)}{f(n_{i+1})}\cr 
- u(n_{i-1},n_i,n_{i+1})=h(n_i,n_{i+1})- h(n_{i-1},n_i) \label{eq:R_1}
\eea
where $h$ is an arbitrary function, yet  to be determined. 
Since the hop rate  $u(n_{i-1},n_i,n_{i+1})=0 $ when  $n_i=0$ and  we must have a boundary condition   $f(m<0)=0,$ 
we can  use    specific values   of $n_i$s  in Eq. (\ref{eq:R_1})  to  find      recursion relation    for $h$. 
For  $n_i=0=n_{i+1}$   equation  (\ref{eq:R_1})    in  $h(n_{i-1},0)=h(0,0).$  Again    putting $n_{i+1}=0=n_{i-1}$ we get
$h(0,n_i)-h(0,0)=u(0,n_i,0).$

These two conditions leaves     Eq. (\ref{eq:R_1})   for   $n_i=0$   as 
\bea
u(n_{i-1},1,n_{i+1}-1)\frac{f(1)}{f(0)}\frac{f(n_{i+1}-1)}{f(n_{i+1})}=u(0,n_{i+1},0). \nonumber 
\eea
Clearly,  in order to be consistent,   $u(n_{i-1},1,n_{i+1})$ must be independent of $n_{i-1}$.  For convenience, 
without any loss of generality,   lets set $u(n_{i-1},1,n_{i+1})=u(0,1,n_{i+1})$.
Thus, to have the factorized steady state for $R=1,$   the hop rate $u(n_{i-1},n_i,n_{i+1})$ must  satisfy 
\bea
u(n_{i-1},n_i+1,n_{i+1}-1) \frac{u(0,1,n_i)}{u(0,n_i+1,0)}  \frac{u(0,n_{i+1},0)}{u(0,1,n_{i+1}-1)} \cr -u(n_{i-1},n_i,n_{i+1})=u(n_{i},n_{i+1},0)-u(n_{i-1},n_i,0). \label{eq:rate_constraint} \nn
\eea
Now if we take   $n_i=1$ and   use  $u(n_{i-1},1,n_{i+1})=u(0,1,n_{i+1})$  in the above equation  to  rearrange the  terms, we have 
\bea
u(n_{i-1},2,n_{i+1}-1) \frac{u(0,1,1)}{u(0,2,0)} \frac{u(0,n_{i+1},0)}{u(0,1,n_{i+1}-1)}\cr  -u(0,1,n_{i+1})= u(1,n_{i+1},0)- u(0,1,0), \nonumber
\eea
which    implies  that  $u(n_{i-1},2,n_{i+1})$ must be independent of $n_{i-1}$.    A similar recursion  would result 
that   $u(n_{i-1},n_i,n_{i+1})$ must be independent of  $n_{i-1}.$  This again reflects the fact that a  factorized steady state is possible 
for $R=1$  only  when  hop rate is  $u=u(n_i,n_{i+1})$  {i.e.}  the process is  a  \textit{misanthrope} process.

\section*{Appendix  B : NO FSS FOR HOP RATE HAVING PRODUCT FORM }

In this section, we show  that the FRP, for any $R>0$, cannot have a FSS when the hop rate has the following product form, 
\be
u(n_{i-R}, \dots, n_i, \dots, n_{i+R})=\prod_{j=-R}^R  v_{j}(n_{i+j}) \label{eq:product_u}.
\ee
The Master equation along with  a  demand  of     a factorized steady state   of the 
form  (\ref{eq:FSS}), and then  Eqs. (\ref{eq:F}) and (\ref{eq:h}) together, implies
\bea
&&v_{-R}\dots v_{-1} v_2 \dots v_R G(n_i,n_{i+1})=h(n_{i-R},\dots,n_{i+R-1})\cr
&&~~~~~~~~~~~~-h(n_{i-R+1},\dots,n_{i+R})  \label{eq:vG}
\eea
where  $v_k \equiv    v_k (n_{i+k})$ and 
\bea
G(n_i,n_{i+1})=&-&v_0(n_i+1)v_1(n_{i+1})\frac{f(n_i+1)f(n_{i+1}-1)}{f(n_i)f(n_{i+1})}\cr
 &+& v_0(n_i)v_1(n_{i+1}).\nonumber 
\eea
Now differentiating both sides of Eq. (\ref{eq:vG})  with respect to $n_{i-R}$ and $n_{i+R}$, we have
\bea
\frac{\partial v_{-R}}{\partial n_{i-R}}\frac{\partial v_{R}}{\partial  n_{i+R}} v_{-R+1}\dots v_{-1}v_2\dots v_{R-1}G(n_i,n_{i+1}) =0 .
\nonumber 
\eea
This implies that, either $v_{-R}(n_{i-R})$  or $v_{R}(n_{i+R})$ must be a constant, 
because   the  other solution $f(n)=1/v_0(n)=1/v_1(n)$   cannot be accepted as it means $v_1(0)=v_0(0)=0$, {\it i.e.}, a 
particle cannot  be  transferred  to a vacant neighbouring site.   So, let us proceed with $v_{-R}=$constant (say $1$). 
Then Eq. (\ref{eq:h}) reads as 
\bea
&&v_{1-R}.. v_{-1} v_2..v_R G(n_i,n_{i+1})=h(n_{i-R},..,n_i,..,n_{i+R-1})\cr
&&~~~~~~~~~~~~~~~-h(n_{i-R+1},..,n_{i},..,n_{i+R}). \nn  
\eea
Clearly for this equation to be valid its right hand side must be independent of $n_{1-R}$ and that in turn  leads to 
$h(x_1,x_2,\dots,x_k)=h(x_2,\dots,x_k)$.

This   way we can eliminate one variable at each step until finally we reach to
\bea
v(n_{i-R}, \dots, n_i, \dots, n_{i+R})=v(n_i,n_{i+1})=v_0(n_i)v_1(n_{i+1}),\cr
{~\rm and ~} v_0(n_i+1)v_1(n_{i+1}-1) \frac{f(n_i+1)}{f(n_i)}\frac{f(n_{i+1}-1)}{f(n_{i+1})}\cr
-v_0(n_i)v_1(n_{i+1})=h(n_{i+1})-h(n_i).\nn
\eea 
This is exactly the   criterion  for   having  a  factorized steady state in  misanthrope process with     a hop rate that  has 
a product form  $u(n_i, n_{i+1}) =  v_0(n_i)  v_1(n_{i+1}) $ \cite{Beyond_ZRP}.

\end{document}